\documentclass[prb,twocolumn,floatfix,showpacs,preprintnumbers, 
               superscriptaddress,amsmath,amssymb]{revtex4}
\usepackage[dvips]{graphics}
\usepackage{psfrag}
\usepackage{graphicx}
\usepackage{epsf}
\usepackage{ucs}
\usepackage[utf8x]{inputenc}
\usepackage{amssymb}
\newcommand{\be}{\begin{eqnarray}}
\newcommand{\ee}{\end{eqnarray}}

\begin{document}
\title{Non-equilibrium polaron hopping transport through DNA}

\author {Benjamin B. Schmidt} 
\affiliation{Institut f\"ur Theoretische Festk\"orperphysik
and DFG-Center for Functional Nanostructures (CFN),
Universit\"at Karlsruhe, 76128 Karlsruhe, Germany}
\affiliation{Forschungszentrum Karlsruhe, Institut f\"ur Nanotechnologie,
  Postfach 3640, 76021 Karlsruhe, Germany}
\author {Matthias H. Hettler} 
\affiliation{Forschungszentrum Karlsruhe, Institut f\"ur
  Nanotechnologie, Postfach 3640, 76021 Karlsruhe, Germany}
\author {Gerd Sch\"on} 
\affiliation{Institut f\"ur Theoretische Festk\"orperphysik
and DFG-Center for Functional Nanostructures (CFN),
Universit\"at Karlsruhe, 76128 Karlsruhe, Germany}
\affiliation{Forschungszentrum Karlsruhe, Institut f\"ur
  Nanotechnologie, Postfach 3640, 76021 Karlsruhe, Germany}

\date{\today}

\begin{abstract}
We study the electronic transport through short DNA chains 
with various sequences of base pairs between voltage-biased leads. 
The strong coupling of the charge carriers to local vibrations 
of the base pairs leads to the formation of polarons, and in the relevant temperature
range the transport is accomplished by sequential polaron hopping.
We calculate the rates for these processes, extending what is known as 
the $P(E)$-theory of single-electron tunneling to the situation with site-specific 
local oscillators. 
The non-equilibrium charge rearrangement along the DNA leads to sequence-dependent current 
thresholds of the `semi-conducting' current-voltage characteristics and,
except for symmetric sequences, to rectifying behavior. The current is thermally activated with activation energy approaching for voltages above the threshold the bulk value 
(polaron shift or reorganization energy). Our results are consistent with some recent experiments.\cite{Yoo01}
\end{abstract}

\maketitle

\section{Introduction} 
\label{sec:Introduction}
Experiments on long-range equilibrium charge transfer along DNA chains indicate
that the dominant transport mechanism is polaron hopping of holes between the 
HOMOs of adjacent base pairs in the DNA stack.\cite{Henderson99,Wan99,Schuster00,Giese00,Behrens02,Joy05} 
In most of these experiments a hole is injected into a guanine donor base by photo-chemical methods. 
The fraction of holes reaching other guanine base traps at a distance,  
separated from the donor guanine by various bridges is recorded. 
In the simplest case the bridges consist of a number of adenine bases, 
but more complicated bridges were investigated as well. 
The experiments showed a weak distance dependence for bridges longer than a few base pairs, which
is in agreement with an activated hopping mechanism. Several theoretical articles argue
that holes are localized on single (guanine) bases either by either solvation effects
and/or structural reorganization.\cite{Voityuk05,Uskov08,Olofsson01,Alexandre03} This localization
can be interpreted as a polaron. The degree of localization is still a matter of debate~\cite{Joy06},
but many authors agree that conformational motion of the DNA is important to charge migration in DNA.\cite{Bruinsma00,Oneill04,Starikov05} For short distances between guanine bases 
(number of intermediate adenine bases \mbox{$N\leq$ 3}) the polaron migrates via superexchange 
(tunneling) between the guanine bases, whereas for large distances the polaron undergoes hopping transport.\cite{Grozema00,Giese01,Berlin01,Berlin02} 
Several authors~\cite{Conwell00,Asai03,Berashevich07} have used an equilibrium polaron hopping picture to model results of various experiments.

On the other hand, in non-equilibrium experiments, where short DNA 
chains were coupled to voltage-biased leads, different types of 
conduction were observed, reaching from quasimetallic~\cite{Xu04} via 
semiconducting~\cite{Porath00,Shigematsu03,Cohen05} to insulating behavior.\cite{Braun98} 
This variance has not been explained (see, e.g., the review by Endres \cite{Endres04}), 
but it suggests a strong influence of the environment and vibrations. 
Furthermore in several experiments~\cite{Yoo01,Roy07} a strong 
temperature dependence consistent with polaron transport was found 
(though another experiment~\cite{Zalinge06} shows only weak temperature dependence).
These observations suggest that polaron hopping plays an important role also
in non-equilibrium transport through DNA. 

In most of earlier work dealing with electronic transport through DNA the
vibrational excitations were either 
neglected altogether, \cite{Roche03,Cuniberti02,Klotsa05} or they were included 
on a simplified level.~\cite{Adessi03,Gutierrez05,Gutierrez06,Schmidt07} 
Other approaches describe polaron hopping as a
classical biased random walk with hopping rates obtained by fits to experiment~\cite{Berlin01,Bixon05} or
quantum chemistry calculations,~\cite{DNAYan2002} 
sometimes including Coulomb charging effects,~\cite{Semrau05} or they treat the 
problem numerically starting from the time-dependent Schr\"odinger equation.\cite{Hultell06}
In this paper, we study non-equilibrium polaron transport in short DNA strands 
connected to voltage-biased leads in the frame of rate equations for polaron hopping, 
with rates accounting for the transitions in the vibrational degrees of freedom as well. 

The DNA is modelled by a tight-binding chain where every 
site corresponds to either a guanine-cytosine (GC) or an adenine-thymine (AT) base-pair, 
which have different on-site energies. Since both DNA strands have a direction (indicated by 5' and 3' at the ends) the local left-right symmetry is broken, and the hopping matrix elements depend not only on the combination of base pairs involved but also on the direction along the DNA molecule 
(compare table~\ref{tab:hopping}).
Every base pair further couples strongly to a local vibration,
thus forming what is known as a `small polaron'. 
Each vibrational degree of freedom, in turn, is coupled to an environment, which is responsible for dissipation and leads to a thermal occupation of the vibrational states.
We will evaluate the rates for polaron hopping in the spirit of what is known 
as the $P(E)$-theory for electron tunneling in a dissipative environment modelled by a bath of oscillators.\cite{Odintsov88,Devoret90,Odintsov91} 
Here, instead of  a bath of oscillators we have for each DNA base pair one localized vibrational mode which, however, is broadened due to the coupling to a dissipative environment. 

Our main results are the following: 
(i) For DNAs with different sequences we observe pronounced differences in the `semiconducting' $I$-$V$ characteristics. 
(ii) For (the usual case of) non-symmetric sequences we observe strong rectifying behavior in the transport due to 
sequence-dependent hopping rates and occupation of the sites.
(iii) For inhomogeneous sequences the current threshold positions 
are not directly connected to intrinsic energy scales, but depend on the charge rearrangement 
at finite bias.
(iv) The current shows a strong temperature dependence following an Arrhenius law. The 
activation energy $E_a$ is voltage dependent and approaches the bulk polaron value $E_a=\Delta/2$  
for high voltages (where $\Delta$ is the polaron shift or reorganization energy).

\section{Model and Technique}
\label{sec:model_and_technique}
We model the DNA with $N$ base pairs by a minimal tight binding model, identifying every base pair with one electronic site. This is motivated by the fact that the molecular orbitals (HOMOs) of the charge carriers (holes) are located mainly on the purine bases, G or A.~\cite{Starikov03,Schmidt07}
We describe the polaron hopping making use of the small-polaron theory, which has been developed to describe strong electron vibration coupling. The important point is that every DNA base can vibrate (quasi) independently from its neighbors, i.e.\ every site is connected to an independent oscillator.
The vibrations in turn are coupled to a dissipative environment which allows for
energy dissipation and relaxation to a thermal occupation. This environment can be modelled by coupling each oscillator to its own bath.

Thus the Hamiltonian is $H=H_{\rm el}+ H_{\rm L}+H_{\rm R}+H_{\rm T,L}+H_{\rm T,R}+H_{\rm vib}+H_{\rm el-vib}+H_{\rm bath}$, with
\begin{eqnarray}
H_{\rm el} &=& \sum_i \epsilon_i a_i^{\dagger}a_i -\sum_{<ij>} t_{ij} a_i^{\dagger}a_j \nonumber\\
H_{\rm T,L}+H_{\rm T,R} &=& \sum_{\nu,r,i} \left[ t^{r}c_{\nu r}^{\dagger}a_i+t^{r*}a_i^{\dagger}c_{\nu r} \right]\nonumber\\
H_{\rm vib}&=& \sum_i \hbar \omega_{i} \left( B_{i}^{\dagger}B_{i} + \frac{1}{2} \right) \nonumber\\
H_{\rm el-vib} &=& \sum_i \lambda_{i} \, a_i^{\dagger}a_i \left(B_{i}+B_{i}^{\dagger}\right)\; .
\end{eqnarray}
The first term $H_{\rm el}$ describes the electrons in the DNA chain with
operators $a_{i}^{\dag}$ and $a_{i}$ in a single-orbital
tight-binding representation with on-site energies 
$\epsilon_i$ of the base pairs and hopping $t_{ij}$ between 
neighboring base pairs.
Both on-site energies and hopping depend on the base pair 
sequence, e.g., the on-site energies differ for Guanine-Cytosine and Adenine-Thymine base pairs.
For the direction-dependent hopping matrix elements $t_{ij}$ we use the values obtained by Siebbeles {\sl et al}.\cite{Senthilkumar05} who studied intra- and 
inter-strand hopping between the bases in DNA-dimers by density functional theory.
Adapting these results to our model of base pairs we obtain the hopping elements 
listed in table~\ref{tab:hopping}, where, e.g., 
the number in the row G and the column A denotes the hopping matrix element
from a GC base pair to an AT base pair to its `right', i.e., in the 3'
direction.\cite{note1}
\begin{table}
5'-XY-3'(all in eV)\\
\begin{tabular}{|c|c|c|c|c|}\hline 
X$\diagdown$ Y & G  & C  & A  & T \\ \hline 
G & 0.119  & 0.046  & -0.186  & -0.048 \\ \hline 
C &  -0.075 & 0.119  & -0.037  & -0.013 \\ \hline 
A & -0.013 & -0.048  & -0.038  & 0.122 \\ \hline 
T & -0.037 & -0.186  & 0.148  & -0.038 \\ \hline 
\end{tabular}
\caption{Hopping integrals $t_{ij}$ taken from Ref.~\onlinecite{Senthilkumar05} 
and adapted to our model. The notation 5'-XY-3' indicates the direction
along the DNA strand (see, e.g., Fig.~1b in Ref.~\onlinecite{Endres04}.) \label{tab:hopping}}
\end{table}

The terms $H_{\rm L,R}$(not written explicitly) with $r=\rm L,R$ refer to the left and right electrodes. 
They are modeled by non-interacting electrons, with operators 
$c_{\nu, \rm r}^{\dag}$ and $c_{\nu\,\rm  r}$,
with a flat density of states $\rho_e$ (wide band limit).
The details of the coupling between the DNA and the electrodes
are not the focus of this work. For our purposes it is sufficiently described by 
$H_{\rm T,L}+H_{\rm T,R}$ with tunneling amplitudes assumed to be independent of the base pair $i$ and the quantum numbers of the electrode states $\nu$. The coupling strength is then characterized by the parameter $\Gamma^{\rm L,R}\propto \rho_e |t^{\rm L,R}|^2$. 

The vibrational degrees of freedom of base $i$ are described by $H_{\rm vib}$, 
with bosonic operators $B_{i}^{\dagger}$ and $B_{i}$ for the mode 
with frequency $\omega_{i}$. The coupling of the electrons
on the DNA to the vibrational modes is described by $H_{\rm el-vib}$, 
where $\lambda_{i}$ is the local electron-vibration coupling strength. 
Here we consider only the so-called stretch modes 
with frequencies $\hbar \omega_i=16\,\rm{meV}$ for a GC base pair 
and $\hbar \omega_i=11\,\rm{meV}$ for an AT base pair, which as shown by  
Starikov couple strongly to the electrons.~\cite{Starikov05}
The coupling strengths are chosen in such a way that the reorganization energy 
or polaron shifts, $\Delta_{\rm A}=0.18\,\rm{eV}$ and $\Delta_{\rm G}=0.47\,\rm{eV}$, 
fit the values extracted from experiments and listed by Olofsson {\sl et al}.\cite{Olofsson01}.
These values probably underestimate the effect of the solvent on the reorganization energy.

The vibration of each base pair $i$ is coupled to the local environment,
$H_{i,\rm bath}$, the 
microscopic details of which do not matter. 
It changes the vibrations' spectra from discrete modes $\omega_i$ to continuous spectra,
\begin{align}
D_i(\omega)=&-i\int dt e^{i \omega t }\theta(t) \left\langle 
\left\lbrace B_i^{\dagger}(t)+B_i(t),B_i^{\dagger}+B_i\right\rbrace   \right\rangle \nonumber\\
=&\frac{1}{\pi}\left(
  \frac{\eta_i(\omega)}{(\omega-\omega_i)^2+\eta_i(\omega)^2}-
\frac{\eta_i(\omega)}{(\omega+\omega_i)^2+\eta_i(\omega)^2}\right)\, .
\end{align}
with frequency dependent broadening $\eta_i(\omega)$.~\cite{Galperin04} The actual form of $\eta_i(\omega)$ depends on the properties of the bath. A reasonable choice which assures also convergence at low and high frequencies 
is $\eta_i(\omega)=\eta_0 \, \frac{\omega^3}{\omega_i^3} \, \theta(\omega_{c}-\omega)$
with $\eta_0=0.5\,{\rm meV}$ and a cutoff of the order of $\hbar \omega_{c}=0.045\,{\rm meV}$.
The coupling to the bath introduces the dissipation, which is crucial for the stability of the DNA molecule in current carrying situations where substantial amount of heat can be produced in the DNA.

In order to describe the system with strong electron-vibration coupling we first apply the so-called polaron or Lang-Firsov unitary transformation 
\be\tilde{H}=e^{S}He^{-S}
\ee
with the generator 
\be
S=-\sum_{i}
\frac{\lambda_{i}}{\hbar \omega_{i}} \, a_i^{\dagger}a_i
\left[B_{i}-B_{i}^{\dagger} \right]\, .
\ee
We introduce transformed electron and vibrational operators,
\be
\tilde{a}_i  &=&  a_i\chi_i \nonumber\\
\tilde{B}_{i} &=& B_{i}-\frac{\lambda_{i}}{\hbar \omega_{i}} \, a_i^{\dagger}a_i \nonumber
\ee
and polaron operators 
\be
\chi_i &=& \exp \left[ \frac{\lambda_{i}}{\hbar \omega_{i}} \,
  (B_{i}-B_{i}^{\dagger}) \right]\, .
\ee
Operators $\chi_i$ with different indices $i$ act on different 
vibrational states, therefore they commute for all times.
In terms of these quantities the Hamiltonian  reads
\begin{align}
\tilde{H} =& \tilde{H}_0 + \tilde{H}'   \\
\tilde{H}_0=& \sum_i (\epsilon_i-\Delta_i) a_i^{\dagger}a_i+  \sum_i \hbar \omega_{i} \left(B_{i}^{\dagger}B_{i}+\frac{1}{2}\right)\nonumber \\
&+H_{\rm L} +H_{\rm R} \\
\tilde{H}'=&-\sum_{<ij>} t_{ij} \, a_i^{\dagger} \chi_i^{\dagger} a_j \chi_j \nonumber \\
&+ \sum_{\nu,r,i} \left[ t^{r} c_{\nu r}^{\dagger} a_i \chi_i +t^{r*}a_i^{\dagger} \chi^{\dagger}_i c_{\nu r} \right] \\
\Delta_i =&\int d\omega D_i(\omega) \frac{\lambda_{i}^2}{\hbar \omega} \;\label{eg:delta} .
\end{align}
After these transformations we can proceed studying the effect of strong electron-vibration coupling
in perturbation theory in $\tilde{H}'$. 
The small parameters are $t_{ij}/\Delta_i$ and $t^{r}/\Delta_i$, which allows 
truncating the perturbation expansion at  lowest non-vanishing order in these parameters.
From here on we will use the shifted on-site energy $\tilde{\epsilon}_i=\epsilon_{i}-\Delta_i$ 
in all expressions.

\subsection*{Rate equation and current}
The small-polaron theory covers two limits of transport. At sufficiently low temperatures  polarons form bands with bandwidth 
$W\simeq W_0\exp\left[-\left(\frac{\lambda}{\hbar \omega}\right)^2 \right]$,
where $W_0$ denotes the electronic bandwidth without vibrations.~\cite{Firsov07} At high temperatures the bandwidth $W$ 
decreases exponentially as the increasing number of multi-phonon processes destroy 
the coherence, and the band picture ceases to be valid. Transport is then accomplished by a sequence of 
incoherent polaron hops. A rough estimate for the cross-over temperature is 
$k_B T\simeq \hbar \omega\left[4\ln\left(\lambda/\hbar \omega\right)\right]^{-1}$.~\cite{Alexandrov95} 
For the electron-vibration coupling strengths of interest in the present problem, room temperature 
is already well above this limit. 

To describe room-temperature transport it is therefore sufficient to consider a rate equation  
for the diagonal elements of the single particle density matrix, 
i.e. the occupation numbers of the sites 
$\rho_l(t)=\left\langle a_l^{\dagger}(t) a_l(t)\right\rangle$.
These occupation numbers evolve according to a master equation with transition rates which we obtain in an expansion in $\tilde{H}'$ from Fermi's Golden Rule.
If we consider the rate for a hopping process from base pair (site) $l$ to $m$, we have to take into account that also the vibrational states may change. If the initial and final states of the coupled system are denoted by $I$ and $F$,
 the rates are
\be
\mathcal{W}_{lm}&=\frac{2 \pi}{\hbar}|t_{lm}|^2 \left| \left\langle F \left|a_m^{\dagger} \chi_m^{\dagger}  a_l \chi_l  \right| I \right\rangle \right|^2 \delta(E_I-E_F).
\ee
In the following the vibrational states are not explicitly considered. 
Therefore, we trace out the vibrational degrees of freedom $X_{l,m}$ by summing over all 
initial vibrational states weighted by the appropriate thermal probability  and over all final state. 
Thus the transition rate from a state with site $l$ initially occupied and site $m$ initially empty becomes
\begin{align}
\mathcal{W}_{lm}=&\frac{2 \pi}{\hbar}|t_{lm}|^2 
 \sum_{X_l,X_l'} \varrho_l(X_l) \left|\left\langle X_l'\left| \chi_l \right| X_l \right\rangle \right|^2\nonumber\\
&\times \sum_{X_m,X_m'} \varrho_m(X_m) \left|\left\langle X_m'\left| \chi_m^{\dagger} \right| X_m \right\rangle \right|^2\nonumber\\
&\times \delta(\tilde{\epsilon}_l-\tilde{\epsilon}_m +E_{X_l}-E_{X_l'}+E_{X_m}-E_{X_m'})\nonumber\;,
\end{align}
where $\varrho_l(X_{l})$ is the probability of finding vibration $l$ in state $X_{l}$. 
Rewriting the energy conserving delta-function by its Fourier transform 
we obtain
\begin{align}
\mathcal{W}_{lm}=&\frac{1}{\hbar^2}|t_{lm}|^2  
\int dt\, e^{\frac{i}{\hbar} \left(\tilde{\epsilon}_l-\tilde{\epsilon}_m \right) t}P_l(t)P_m(t)
  \,, \label{eq:hopW1}
\end{align}
where
\begin{align}
P_l(t)=&\sum_{X_l} \varrho_l(X_l) \left\langle X_l \left| \chi_l^{\dagger}(t) \chi_l(0) \right| X_l \right\rangle\\
=&\mathcal{K}_{l} \exp\left[ \int d\omega D_l(\omega) \left(\frac{\lambda_{l}}{\hbar\omega} \right)^2 \frac{\cos\left(\omega \left[t+i\hbar\beta/2 \right]\right)}{\sinh\left(\hbar \omega\beta/2 \right)}  \right]\; ,\nonumber 
\end{align}
with
\begin{align}
\mathcal{K}_{l}=&\exp \left\lbrace -\int d\omega D_{l}(\omega)
\left(\frac{\lambda_{l}}{\hbar\omega}\right)^2  \coth\left(\hbar \omega\beta/2 \right) \right\rbrace \; .
\end{align}
The function $P_l(t)$ is known from the ``$P(E)$ theory'', which describes tunneling in a dissipative electromagnetic environment, modelled by an infinite set of oscillators. Here, instead of such a bath we have broadened local vibrational modes of two DNA base pairs involved in the hopping process.

The calculation for the tunneling transition between the left ($L$) 
and right ($R$) electrodes and the first or last site of the DNA chain $l=1$ or $l=N$ proceeds similarly, except that one has to trace also over the electrodes' electronic states, while we have to consider only the local vibration of the one site involved. Hence we have for the rates on the left, out and onto the DNA chain
\begin{align}
W^{\rm L}_-=&\Gamma^{\rm L}  \int \frac{dE}{2\pi\hbar} (1-f_{\rm L}(E)) P_1(\tilde{\epsilon}_{1}-E)\nonumber \\
W^{\rm L}_+=&\Gamma^{\rm L} \int \frac{dE}{2\pi\hbar} f_{\rm L}(E) P_1(E-\tilde{\epsilon}_{1})\label{eq:tunVL}\; ,
\end{align} 
where $\Gamma^{\rm L} = 2\pi |t^{\rm L}|^2 \rho_e$, 
$f_{\rm L}(E)$ is the Fermi function in left lead, and  $P_1(E)$ is the Fourier transform of $P_1(t)$.  For the right interface a similar expression holds involving $f_{\rm R}(E)$ and $P_N(E)$.

The master equation for sites in the DNA chain thus reads
\begin{align}
\frac{d}{dt}\rho_{l}=& \sum_{m} \Big[ - \rho_{l}\left(1-\rho_{m}\right)\mathcal{W}_{l m}+ \left(1-\rho_{l}\right)\rho_{m}\mathcal{W}_{m l} \Big] \label{eq:dt_density1},
\end{align}
where the sum over $m$ is restricted to nearest neighbors of $l$.\cite{note3}
For the base pair at the left end of the chain we get
\begin{align}
\frac{d}{dt}\rho_{1}=& -\rho_{1} W^{\rm L}_-+\left(1-\rho_{1}\right) W^{\rm L}_+\nonumber\\
&+ \Big[ -\rho_{1}\left(1-\rho_{2}\right)\mathcal{W}_{1 2}+ \left(1-\rho_{1}\right)\rho_{2}\mathcal{W}_{2
1} \Big]\; ,\label{eq:dt_density3}
\end{align}
and similar for the right interface.

We are interested in the steady state, $d\rho_{l}/dt=0$, which develops for a constant applied bias. After solving the resulting self-consistent equations  iteratively 
we can calculate the non-equilibrium current through the left lead,
\begin{align}
I_{\rm L}=e  \Big[& -\rho_{1} W^{\rm L}_- +\left(1-\rho_{1}\right) W^{\rm L}_+ \Big]
\end{align}
or for the right lead, which is the same since the current is 
conserved, $I_{\rm L}=-I_{\rm R}$.

\subsection*{Discussion of the hopping rates}
For the hopping rates eq.~(\ref{eq:hopW1}) the situation differs from the usual $P(E)$ theory: instead of one infinite vibrational bath 
each base pair ($m$ and $l$) has its own vibration degree of freedom and  we get products $P_{m}(t) P_{l}(t)$, which become convolutions in energy space. The rates still satisfy detailed balance
\begin{align}
\mathcal{W}_{l m}=\mathcal{W}_{m l} \exp\left[\frac{\tilde{\epsilon}_{l}-\tilde{\epsilon}_{m}}{k_{\rm B} T} \right]\;,\label{eq:detbal}
\end{align}
where $\tilde{\epsilon}_{m}$ and $\tilde{\epsilon}_{l}$ are the on-site energies of base pairs $m$ and 
$l$, respectively. 

For large times, $P_l(t)$ approaches a constant, $\mathop {\lim }\limits_{t \to \infty } P_l(t)=\mathcal{K}_{l}$. 
Therefore it can  be separated into two terms, 
one decaying in time and one constant:
\begin{align}
P_l(t)=\tilde{P_l}(t)+\mathcal{K}_{l}
\label{eq:sep}\;.
\end{align}
Accordingly we can write
\begin{align}
P_{l}(t)P_{m}(t)
=\tilde{P}_l(t)\tilde{P}_m(t)+\mathcal{K}_{m}\tilde{P}_l(t)+\mathcal{K}_{l}\tilde{P}_m(t)+\mathcal{K}_{m}\mathcal{K}_{l}\nonumber\,.
\end{align}
The product $\tilde{P}_l(t)\tilde{P}_m(t)$ describe transitions, where the number of vibrations changes on both sites, the next two terms
describe changes in one of the two sites only, while the last term describes transitions without changes in the vibration state.
When performing the time integration in eq.~(\ref{eq:hopW1}), this last term
leads to a divergence when the two site energies are degenerate
\begin{align}
\frac{1}{2 \pi \hbar}\int dt \, e^{\frac{i}{\hbar}\left(\tilde{\epsilon}_{l}-\tilde{\epsilon}_{m}\right)t} \mathcal{K}_{m}\mathcal{K}_{l}=
\mathcal{K}_{l}\mathcal{K}_{m}\delta\left(\tilde{\epsilon}_{l}-\tilde{\epsilon}_{m}\right)\,, 
\label{eq:subtraction}
\end{align}
since in this situation the phenomena of resonant tunneling occurs. 
In this situation the perturbation theory limited to second order is not sufficient. Rather, one should sum up in a `ladder'-approximation
an infinite series of such terms, leading to a result with finite rates.~\cite{Boettger85,Boettger93,Konstantinov61}

Alternatively, we can phenomenologically regularize the divergence of 
eq.~(\ref{eq:subtraction}) by formally introducing an imaginary part 
to the level energies $\tilde{\epsilon}_{l}$. This is motivated by the fact that they aquire a finite width due to the interaction with the vibrations or leads. In this way the hopping rates become finite. We further note, that the contribution of the vibration-free transitions (the constant term of eq.~\ref{eq:subtraction}) is multiplied by the factor $\mathcal{K}_{l}\mathcal{K}_{m}$. 
In our case, this factor is exponentially small. This corresponds to the fact that we consider the limit where polaron hopping by far dominates polaron band transport.
We therefore can ignore the terms $\propto \mathcal{K}_{m}\mathcal{K}_{l}$ in our analysis all together, i.e.\ we subtract them in eq.~(\ref{eq:hopW1}). The regularized hopping rates are  therefore
\begin{align}
\mathcal{W}_{l m}=&\frac{|t_{l,m}|^2}{\hbar^2} \int dt \, e^{\frac{i}{\hbar}\left(\tilde{\epsilon}_{l}-\tilde{\epsilon}_{m}\right)t}\, \Big[P_{l}(t) P_{m}(t)-\mathcal{K}_{l}\mathcal{K}_{m}\Big] \; .\label{eq:hopW}
\end{align}

\section{Results}
\subsection*{Sequence effects}
\label{sec:results}
Using the rate equation we will study now the charge transport  and the non-equilibrium occupation of the sites for various DNA sequences. Both quantities 
depend strongly on the specific sequence. All DNA sequences
are `semi-conducting'  since the Fermi energy lies in the HOMO-LUMO gap, i.e. well above the HOMO states which carry the transport.

\begin{figure}
\begin{center}
\includegraphics[width=8cm]{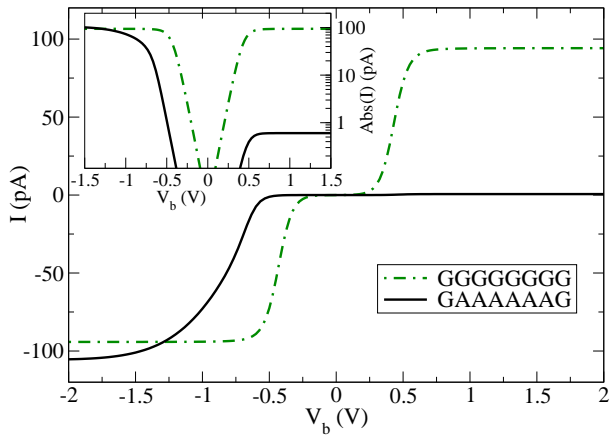}\\
\caption{(Color online) $I$-$V$ characteristics for two DNA strands with sequences 
 5'-GGGGGGGG-3' (dash-dotted line) and 5'-GAAAAAAG-3' (solid line)
with the following parameters: Base pair on-site energies 
$\epsilon_{\rm A}=-0.26\,\rm{eV}$, $\epsilon_{\rm G}=+0.25\,\rm{eV}$, polaron shifts 
$\Delta_{\rm A}=0.18\,\rm{eV}$ and $\Delta_{\rm G}=0.47\,\rm{eV}$, Fermi energy $E_{\rm F}=0\,\rm{eV}$,
symmetric coupling to leads with linewidths $\Gamma_{\rm L}=\Gamma_{\rm R}=0.01\,\rm{eV}$,
vibrational energies $\hbar \omega_{\rm A}=11\,\rm{meV}$, $\hbar \omega_{\rm G}=16\,\rm{meV}$, and
room temperature $k_{\rm B}T=25\,\rm{meV}$. The inset shows the absolute value of the current
on logarithmic scale. The current for the second sequence shows  rectification by a factor of $\sim 200$.}\label{fig:IV400}
\end{center}
\end{figure}

Figure~\ref{fig:IV400} shows the $I$-$V$ characteristics for two such sequences, 5'-GGGGGGGG-3' (green, dash-dotted line) and 5'-GAAAAAAG-3' (black, solid line). 
The first sequence displays the `semi-conducting' behavior with a gap characterized 
by the distance of the Fermi energy to the onsite energy of the G base (shifted by $\Delta_{\rm G}$).
Due to its electronic symmetry the $I$-$V$ characteristic is symmetric 
with respect to the applied bias. On the other hand, the second DNA sequence 
shows strong rectifying behavior, despite of its seemingly symmetric sequence. The reason for this asymmetry lies in the electronic asymmetry of the hopping amplitudes, together with the incoherence of the hopping processes between DNA base pairs. This can easily be understood: For positive bias the hopping `bottleneck' 
of the system is at the crossover from 
A to G at the 3' end of the strand. There, the polaron needs to overcome an energy barrier mediated by vibrational excitations. For negative bias the `bottleneck' is at the 
crossover from A to G at the 5' end of the strand.
Due to the opposite direction of the dominating hopping process,
with $|t_{\rm GA}| > |t_{\rm AG}|$ (compare Table~\ref{tab:hopping}),
the current for negative bias is higher than for positive bias. 
Thus, inhomogeneous sequences will in general display a rectifying, semi-conducting $I$-$V$ characteristic.
The rectification effect will be weaker for longer and more disordered sequences, as more `bottlenecks' in 
either direction appear. Note that no rectifying behavior would be observed if we model the transport as a coherent transition through the total length of the chain ( `Landauer approach').~\cite{Gutierrez05,Gutierrez06,Schmidt07}

\begin{figure}
\begin{center}
\includegraphics[width=8cm]{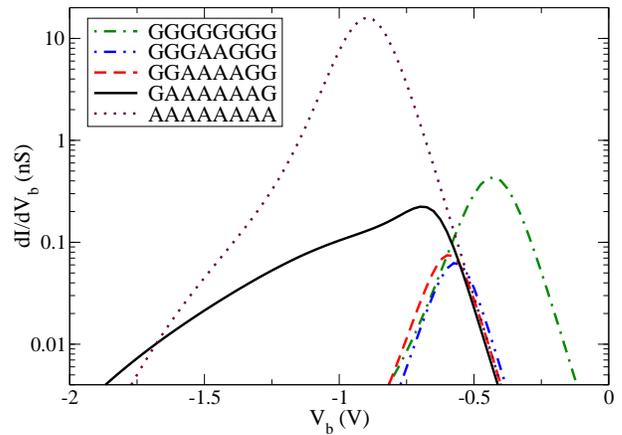}\\
\caption{(Color online) Differential conductance (logarithmic scale) as a function of applied
bias for five different DNA sequences with parameters as in Fig.~\ref{fig:IV400}.
For the homogeneous sequences the threshold, i.e. the position of the maximum 
of the differential conductance, 
is set by on-site energy of the considered base pairs. 
For the inhomogeneous sequences, however, the threshold is not determined by the internal energy scales 
alone. The sequence-dependent thresholds lie in between the limits set by the homogeneous sequences.
For some sequences (e.g. 5'-GAAAAAAG-3'), the peaks are broadened due to `renormalized' tunneling.}
\label{fig:thresh}
\end{center}
\end{figure}

We now study the sequence dependence of the current threshold, or equivalently the postion of the peak in 
the differential conductance $dI/dV_{\rm b}$ (both differ only by a term proportional to temperature). 
Figure~\ref{fig:thresh} shows the differential conductance as a function of the applied
bias for 5 different DNA sequences 5'-AAAAAAAA-3', 5'-GAAAAAAG-3', 5'-GGAAAAGG-3', 
5'-GGGAAGGG-3', and 5'-GGGGGGGG-3'. For the homogeneous sequences the threshold is equal to the 
on-site energy of the considered base pairs
($eV_{\rm b}=2\tilde{\epsilon}_{\rm A}$ for 5'-AAAAAAAA-3' and $eV_{\rm b}=2\tilde{\epsilon}_{\rm G}$ for 5'-GGGGGGGG-3'). 
For the inhomogeneous sequences the threshold lies in between the limits set by the homogeneous 
sequences, i.e. it is not determined by the internal energy scales alone. The varying threshold 
is a consequence of the way the charges are rearranged along the DNA molecule, which of course is very sensitive to the considered sequence.

As discussed above, eqs.~(\ref{eq:tunVL}) describe the tunneling rate from the electrode
to the adjacent DNA base pair. The tunneling is modified by the vibrational modes 
which can be excited, depending on the applied bias. This `renormalized' tunneling
can lead to a very broad differential conductance peak which is very different 
from the usual (derivative of) Fermi function form, as observed most prominently 
for the sequence 5'-GAAAAAAG-3'. Note that there is nearly no modification on the low bias
side of the peak.

\subsection*{Local chemical potential}

As discussed above, the $I$-$V$ characteristic of a DNA molecule is 
affected by bias and sequence dependent charge rearrangements on the DNA base pairs. For the ease of displaying these effects, i.e, both small deviations from an occupation 1, as well as occupations near 0, we introduce a local chemical potential $\Phi_i$, defined by
\begin{align}
\Phi_i(V_{\rm b})=\tilde{\epsilon}_i-k_{\rm B}T \ln\left(\frac{1}{\rho_i(V_{\rm b})}-1\right).\label{eq:phi}
\end{align}
This quantity is superior to the occupation in visualizing the non-equilibrium charge 
rearrangement, because it reacts sensitively to even small changes in the occupation.

\begin{figure}
\begin{center}
\includegraphics[width=8cm]{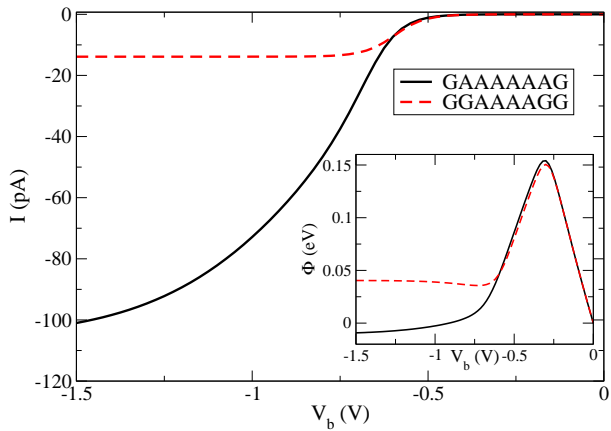}\\
\caption{(Color online) $I$-$V$ curves for the two sequences 5'-GAAAAAAG-3' (solid line) and 
5'-GGAAAAGG-3' (dashed line) (parameters see Fig.~\ref{fig:IV400}). Despite the very 
similar sequences, the $I$-$V$ show clear differences. The inset shows the local chemical 
potential $\Phi$ for the last guanine base (at the 3' end) for both sequences at various 
bias voltages. Equivalent behavior between local potential and $I$-$V$ is visible}\label{fig:ivmu100}
\end{center}
\end{figure}

Figure~\ref{fig:ivmu100} shows the $I$-$V$ curves for the 
two DNA molecules 5'-GAAAAAAG-3' (black, solid line) and 5'-GGAAAAGG-3' (red, dashed line), 
and the inset shows the local chemical potential $\Phi$ for the last guanine base 
(at the 3' end) for both sequences.
Although the sequences are very similar, the $I$-$V$ characteristics differ strongly in the maximum 
current and in the way the current increases for increasing bias voltage. The current of the second
sequence has reached a plateau already at about $V_{\rm b}=-0.8\,\rm{V}$, whereas the black curve has not 
leveled off even for $V_{\rm b}=-1.5V$. This strong deviation from a Fermi function behavior 
is in part a consequence of the renormalization of the tunneling rates by the vibrations.

This difference in the $I$-$V$ characteristics is reflected in the local chemical potential $\Phi$, 
most prominently at the last guanine base of both sequences, as shown in the inset. At low bias both sequences behave in the same way: the potential increases equally with the applied bias.
The DNA is not conducting and therefore, the situation is similar to the charging of a 
capacitor. At the drop-off around $V_{\rm b}=-0.3\,\rm{V}$, the current sets in and a potential drop between base pair 
and lead is established. In correspondence to the current, the local chemical potential 
for the second sequence 5'-GGAAAAGG-3' levels off, whereas the potential of the first 
sequence 5'-GAAAAAAG-3' never reaches a plateau in the range up to $V_{\rm b}=-1.5\,\rm{V}$.

\begin{figure}
\begin{center}
\includegraphics[width=9.5cm]{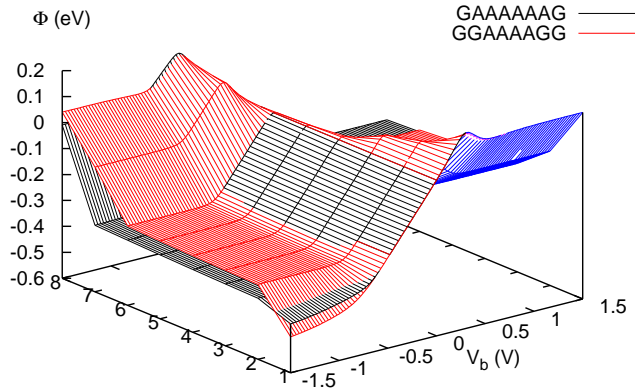}\\
\caption{(Color online) Local chemical potential $\Phi$ for all base pairs of the DNA strand with 
sequence 5'-GAAAAAAG-3' (black lines) and 5'-GGAAAAGG-3' (red lines) at various bias voltages 
and with parameters as in Fig.~\ref{fig:IV400}. The local potential drops differently for the
two sequences, implying different conduction properties. 
}
\label{fig:mu}
\end{center}
\end{figure}

To give a feeling for the total charge rearrangement Figure~\ref{fig:mu} shows the 
local chemical potential $\Phi_i$ of the two DNA sequences for all base pairs $i$ and
all voltages.
The chemical potential landscape also suggests how the bias voltage $V_{\rm b}$ applied to the leads drops
over the entire DNA molecule. Regions of good conductivity show almost no voltage drop, as seen
for the stretches of adenine bases in the middle of both sequences. On the other hand most of the 
voltages drops at the base pairs close to the interfaces. The potential/voltage drop over for the
entire sequence 5'-GAAAAAAG-3' is less than for 5'-GGAAAAGG-3'. This suggests that the the first 
sequence is better conducting than the latter, which is in accordance with their $I$-$V$ curves
(Fig.~\ref{fig:ivmu100}).

\subsection*{Temperature dependence and activation energy}

In the experiments of Ref.~\onlinecite{Yoo01} the current through bundles of long homogeneous DNA molecules showed a strong temperature dependence. The data could be fitted by an activation
law $I(V) = \alpha(V) \exp \left[\frac{-E_a}{k_{\rm B} T} \right]$, with a voltage dependent prefactor $\alpha(V)$ that also shows a temperature dependence for the case poly(dG)-poly(dC) bundles. 
Asai~\cite{Asai03} has used the Kubo formula for a polaron hopping model to obtain a similar relation for the linear response conductivity.

\begin{figure}
\begin{center}
\includegraphics[width=8cm]{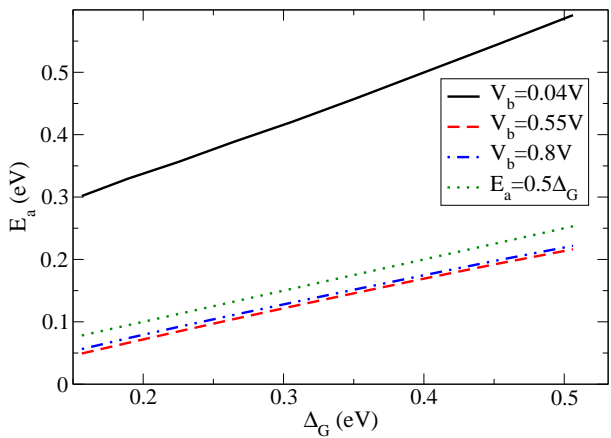}\\
\caption{(Color online) Activation energy $E_a$ for polaron hopping of a homogeneous DNA
with 15 G-C base pairs as a 
function of the polaron shift $\Delta$ for voltages $V_{\rm b}=0.04\,\rm{V}$ (solid line),
$V_{\rm b}=0.55\,\rm{V}$ (dashed line), and $V_{\rm b}=0.8\,\rm{V}$ (dash-dotted line). 
All other parameters as in Fig.~\ref{fig:IV400}.
For comparison, the dotted line shows the activation energy of polaron hopping conduction
in bulk at high temperatures ($E_a=0.5\Delta$).\cite{Boettger85}}\label{fig:activen}
\end{center}
\end{figure}

Our results are obtained in a non-equilibrium situation and also show
a strong temperature dependence. An Arrhenius plot of the current vs. temperature shows linear behavior, indicating that the current is indeed an activated quantity (though we also observe deviations from a perfect Arrhenius law). Fitting the temperature 
dependence of our data by an Arrhenius law allows us to estimate the activation energy for a given bias voltage and polaron shift $\Delta$.
Figure~\ref{fig:activen} shows the activation energy $E_a$  obtained by this fitting
as a function of the polaron shift at three different bias voltages 
for a homogeneous DNA strand with 15 GC base pairs.~\cite{note2}
The activation energy is proportional to $\Delta_{\rm G}$, but the
proportionality factor differs depending on the applied bias voltages. For voltages smaller
than the gap, the activation energy also includes the energy needed to 
overcome the gap. 
For voltages above the threshold the proportionality factor between activation energy and polaron shift
is about $1/2$, consistent with the high-temperature value for the bulk polaron hopping conduction
$E_a=0.5\Delta$ (see green/dotted line in Fig.~\ref{fig:activen}).\cite{Boettger85} 

\section{Summary}
\label{sec:summary}
We have investigated the non-equilibrium polaron hopping transport in short DNA chains with various sequences, 
coupled to voltage-biased leads in the frame of rate equations which take into account inelastic 
transitions in the local vibration degrees of freedom.
Our theory is formally an extension of the so-called $P(E)$ theory
of tunneling in a dissipative electromagnetic environment. 
We find semi-conducting $I$-$V$ characteristics with thresholds that 
are very sensitive to the considered DNA sequence. 
For all non-symmetric sequences (which is the typical case) we observe rectifying behavior (Fig.~\ref{fig:IV400}).
The sequence dependent thresholds are not directly connected to intrinsic energy scales (Fig.~\ref{fig:thresh}), rather they are intimately related to the non-trivial charge rearrangement along the DNA molecule at finite bias.
We have visualized this effect by displaying the local chemical potential $\Phi_i$ (Fig.~\ref{fig:mu}).
As expected for polaron hopping, the current is thermally activated with a temperature dependence following  
an Arrhenius-law. The activation energy $E_a$ is voltage dependent and approaches
the bulk polaron value $E_a=1/2\Delta$ ($\Delta$: polaron shift) for voltages 
above the threshold (Fig.~\ref{fig:activen}).

{\em Acknowledgments.} We acknowledge stimulating discussions with Janne Viljas, 
Elke Scheer, Jewgeni Starikow, and Wolfgang Wenzel. We also thank the  \mbox{Landesstiftung} Baden-W\"urttemberg for
financial support via the \mbox{Kompetenznetz} ``Funktionelle
Nanostrukturen''.


\begin{thebibliography}{49}
\bibitem{note1} We assume that the
  holes can only reside on the purine bases, G or A. 
The hopping integrals between two purines depend
on the specific bases involved and to which strands these two bases
belong to. We use the values for the hopping integral
$J$ of the first, second and fifth row in the table 3 of Ref.~\onlinecite{Senthilkumar05}.
They are reproduced in our table 1.
\bibitem{note3}  Including also hopping to more distant bases, one could account for
the superexchange mechanism, dominant for guanine bases closer than three base pairs, 
as considered in Ref.~\onlinecite{Berlin01}. This is not important for the sequences considered 
in this work, so it was not included in the numerical evaluation.
\bibitem{note2} We chose to set the polaron shift in the electronic part of the Hamiltonian to 
zero, so that the gap between on-site energy $\tilde{\epsilon}_{\rm G}$ 
and the Fermi level is constant for all 
electron-vibration coupling strengths ($|\tilde{\epsilon}_{\rm G}-E_{\rm F}|=0.25eV$). This allows 
us to keep the bias voltages fixed for the range of  polaron shifts considered.
\end{thebibliography}
\end{document}